\documentclass[
  twocolumn,
  superscriptaddress,
  pra,
  amsmath,
  amssymb,
  showkeys
]{revtex4-2}

\usepackage{graphicx}
\usepackage{bm}
\usepackage{color}
\usepackage{hyperref}
\hypersetup{colorlinks,allcolors=blue}
\usepackage[utf8]{inputenc}
\usepackage[T1]{fontenc}

\newcommand{\be}{\begin{equation}}
\newcommand{\ee}{\end{equation}}
\newcommand{\bea}{\begin{eqnarray}}
\newcommand{\eea}{\end{eqnarray}}
\newcommand{\ba}[1]{\begin{array}{#1}}
\newcommand{\ea}{\end{array}}

\begin{document}

\title{On Sub-Sevenfold Symmetries in LH2 Stacked Ring Scaffolds: A Quantum Optical Perspective}

\author{Arpita Pal}
\email{Arpita.Pal@uibk.ac.at}
\affiliation{Institut f\"ur Theoretische Physik, Universit\"at Innsbruck, Technikerstra{\ss}e 21a, A-6020 Innsbruck, Austria}

\begin{abstract}
For highly efficient light-harvesting, purple photosynthetic bacteria rely on light-harvesting 2 (LH2) complexes featuring ninefold, eightfold, and sevenfold rotational symmetries; however, sub-sevenfold scaffolds have not yet been observed in nature. To investigate this absence, we utilize a closed coupled-dipole model that neglects dissipative effects, which remains valid for a temporal window shorter than the interlayer excitation energy transfer time (on the picosecond scale) in an LH2 scaffold. We examine collective eigenmodes across $C_n$ stacked model ring geometries for $n=9,8,7,6$. Our results demonstrate that the minimum energy separation between specific collective bands occurs at the angular-momentum modes $m=\pm 1$ for known geometries, effectively predicting the photoactive absorption wavelengths of the lower bacteriochlorophyll (BChl) ring. Specifically, we predict peaks at 851, 848, and 825 nm for ninefold, eightfold, and sevenfold symmetries, respectively, which agree well with experimental observations. Applying this analysis to a hypothetical sixfold geometry, we predict a significant blue shift in the photoactive wavelength of the lower BChl ring, suggesting inefficient harvesting of the available solar spectrum and offering a quantum optical perspective on the evolutionary absence of sub-sevenfold scaffolds in LH2 complexes. Our study demonstrates that quantum cooperative models can elucidate specific aspects of biological LH systems.
\end{abstract}

\maketitle

\section{Introduction}
\label{sec:intro}
Patterns are apparent everywhere around us~\cite{pickover1995pattern,ball2009shapes,ball2016patterns}, stretching from the largest cosmic scales to the smallest molecular ones. We observe them, for example, in spiral galaxies~\cite{sellwood:araa:2022}, across vast landscapes~\cite{bell:book}, within dynamic auroras~\cite{falck1999aurora} and clouds~\cite{perkins:cloud:2018}, in flowing rivers and windswept desert sands~\cite{ball2011flow}, and in the crystalline lattices of minerals and snowflakes~\cite{tilley2006crystals, libbrecht2015snowflake}. These patterns are equally ubiquitous in living entities, ranging from the macroscopic branching architecture of trees~\cite{ball2009branches} to microscopic biological structures~\cite{blankenship2002molecular,levitan2015neuron}. As \textit{Homo sapiens}, we are naturally driven to ask: why does Nature prefer specific patterns or symmetries? In most cases, these forms emerge not as aesthetic features, but as direct consequences of underlying physical laws or evolutionary constraints~\cite{rosen2008symmetry}. Often, a pattern reflects the most efficient configuration for a specific process - whether by minimizing energy, distributing forces uniformly, or optimizing transport and growth. In this sense, Nature's widespread use of patterns or symmetry is neither coincidental nor decorative; rather, it represents the optimal solution toward which physical, chemical, and biological systems converge to meet the demands of governing principles and achieve maximal efficiency.

Here, we narrow this broad landscape of questions toward a specific and unresolved problem: the absence of sub-sevenfold rotational symmetry within the biological geometry of the light-harvesting (LH) 2 complex in purple photosynthetic bacteria~\cite{McDermott:1995, kuhlbrant:structure:1995, koepke:structure:1996, Sturgis:pr:2009}. This pigment--protein complex is central to the remarkable efficiency of bacterial photosynthesis~\cite{blankenship2002molecular, croce2018light}. Consequently, it has remained a subject of longstanding curiosity, particularly regarding the quantum-chemical principles that enable such efficient light capture and energy transfer~\cite{VANGRONDELLE19941, renger:pccp:2013, scholes:cr:2017, menucci:cr:2017}. To function as an effective antenna and facilitate inter-layer excitation energy transfer (EET), nature relies most abundantly on ninefold oligomeric symmetry, as seen in the LH2 of \textit{Rhodoblastus acidophilus}~\cite{McDermott:1995, kohler_2006}. Additionally, sevenfold symmetry has been identified in \textit{Marichromatium purpuratum}~\cite{qian:sa:2021, cupellini:pr:2023} and eightfold symmetry in \textit{Rhodospirillum molischianum}~\cite{koepke:structure:1996, schulten:pnas:1998}. However, to the best of our knowledge, sub-sevenfold oligomeric symmetries have not been reported for single LH2 ring scaffolds. We therefore ask: why are sub-sevenfold rotational symmetries absent within these scaffolds when there are no apparent structural constraints limiting their occurrence? Furthermore, what underlying physical principles account for this absence, and what logic governs this geometric constraint?

To address these questions, we investigate using theoretical tools from quantum optics and the physics of collective light--matter interactions~\cite{genes:prx:2022, janne:pra:2023}. By analyzing the interplay between the geometric arrangement of bacteriochlorophylls (BChls)~\cite{McDermott:1995, schulten:pnas:1998, montemayor:jpcb:2018}, the symmetry-dependent collective eigenenergies of model LH2 ring geometries, and the local photonic environment of purple bacteria, we illustrate how a quantum-optical perspective of collective light--matter interaction helps to elucidate the absence of sub-sevenfold rotational symmetries in single biological LH2 ring scaffolds. While our findings offer a window into the quantum-optical constraints of these rings, they represent only a fraction of the broad design principles nature employs for light-harvesting systems. This work aims to provide a quantum-optical perspective on the absent sub-sevenfold symmetries in LH2 complexes, complementing existing quantum-chemical descriptions with an approach grounded in collective light--matter interactions. Furthermore, our choice to utilize a `closed coupled-dipole model' (neglecting dissipative influences under reasonable approximations) is an entirely distinct effort from our previous work~\cite{pal:njp:2025}, as detailed in the following sections.

The work is organized as follows. In Sec.~\ref{theory}, we provide the theoretical description of a model stacked ring geometry of deep-subwavelength spaced interacting quantum emitters. In Sec.~\ref{ring-geometries}, we utilize a closed coupled-dipole model without dissipation, valid for a temporal window shorter than interlayer EET, in order to predict the absorption peaks of the bottom ring of LH2 model stacked geometries with ninefold, eightfold, and sevenfold symmetries, which agree well with experimental observations. In Sec.~\ref{sub-sevenfold}, we introduce a hypothetical sixfold stacked ring scaffold and calculate the predicted absorption peak for the bottom ring, accounting for the local photonic environment of the purple bacteria. We argue that these sub-sevenfold geometries are inefficient for light-harvesting, explaining their absence in nature's engineering of the LH2 complex. Finally, Sec.~\ref{conclusion} provides the conclusions and outlook of our study.

\section{Theoretical Description}
\label{theory}
\begin{figure}
\centering
\includegraphics[width=0.9\columnwidth]{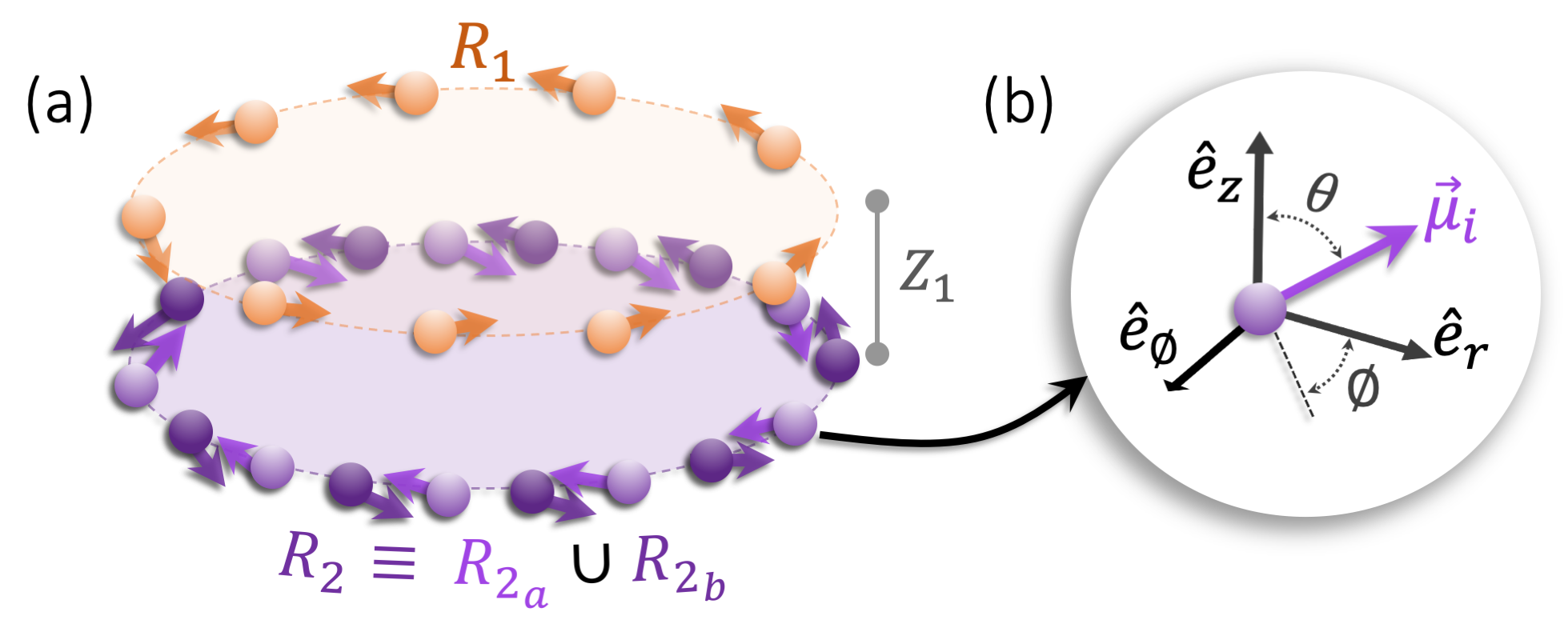}
\caption{(a) Schematic representation of the stacked LH2 scaffold with $C_9$ symmetry, showing the top ring $R_1$ (orange) and the bottom ring $R_2$ (purple), composed of two concentric sub-rings $R_{2_a}$ (light-purple) and $R_{2_b}$ (dark-purple) of deep-subwavelength-spaced atoms. Arrows indicate representative dipole orientations of each atom. The vertical layer separation is $Z_1$. For a more detailed view of the LH2 scaffold with BChls, see Ref.~\cite{kohler_2006}. (b) Local coordinate system for the $i^{th}$ dipole: $\hat{e}_r$, $\hat{e}_\phi$, and $\hat{e}_z$ are the radial, azimuthal, and axial unit vectors, respectively, and the angles $\theta$ and $\phi$ define the orientation of the dipole moment $\vec{\mu}_i$ (adapted from Ref.~\cite{pal:njp:2025}).}
\label{schematic}
\end{figure}

We theoretically consider a very deep subwavelength-spaced atomic ring geometry of interacting dipoles in a three-dimensional (3D) stacked configuration (for visual illustration see Fig.~\ref{schematic}). In this geometry, the top ring $R_1$ has a radius of $r_1$ and contains $N_1$ atoms. The bottom ring, $R_2$, is composed of two concentric sub-rings $R_{2_a}$ and $R_{2_b}$, having radii $r_{2_a}$ and $r_{2_b}$, containing $N_{2_a}$ and $N_{2_b}$ atoms, respectively (with $N_2 = N_{2_a} + N_{2_b}$). The total Hamiltonian for this system of interacting dipoles in a stacked ring geometry is given by
\bea
H_{\rm tot} = \sum_{k\in \{R_1,R_2\}} \omega_{k} \hat{\sigma}^+_k \hat{\sigma}^-_k +H_{\rm DD}~,
\label{H_tot}
\eea
where $\hat{\sigma}^+_i(\hat{\sigma}^-_i)$ are the atomic raising (lowering) operators and $\omega_k = \omega_{\nu_0}$ for $k\in R_1$ and $\omega_k = \omega_{\nu}$ for $k\in R_2$. The atomic transition frequencies $\omega_{\nu_0} (\omega_{\nu})$ correspond to the $\lambda_0(\lambda)$~nm dipoles in rings $R_1$ and $R_2$, respectively. $H_{\rm DD}$ represents the dipole-dipole interaction Hamiltonian, which is given by
\bea
H_{\rm DD} = \sum_{i,j \in S, i\neq j}\Omega_{ij}\,\hat{\sigma}^+_i \hat{\sigma}^-_j~,
\label{H_DD}
\eea
where $S = R_1 \cup R_2$. The collective dipole-dipole coupling between the $i^{th}$ and $j^{th}$ dipoles is given by the expression~\cite{FICEK:pr:2002}
\bea
\Omega_{ij} &=& \frac{3\Gamma_0}{4} \Bigg[(1- 3 \cos^2 \theta_{ij}) \left( \frac{\sin \xi_{ij}}{\xi^2_{ij}} + \frac{\cos\xi_{ij}}{\xi^3_{ij}}\right) \nonumber\\
&& - \sin^2 \theta_{ij} \frac{\sin\xi_{ij}}{\xi_{ij}} \Bigg]~,
\label{omegaij}
\eea
where $\xi_{ij} = k_0 r_{ij}$ represents the dimensionless distance scaled by the wavenumber $k_0 = 2\pi/\lambda_0$ of the transition at wavelength $\lambda_0$ and $\theta_{ij}$ is the angle between dipole moment $\mu = |\vec{\mu}_{i(j)}|$ and the separation vector $\vec{r}_{ij}$. The spontaneous emission rate of a single quantum emitter is $\Gamma_0 = \mu^2 k^3_0/(3\pi\hbar\epsilon_0)$. On-site collective energy shifts are not considered, i.e., $\Omega_{ii} = 0$. Utilizing the collective Bloch eigenmode description with a specific angular momentum quantum number $m$~\cite{Mariona:pra:2019}, the Hamiltonian for a stacked ring configuration with $N_1$-fold symmetry can be re-expressed as follows~\cite{scheil:nm:2023, pal:njp:2025}

\bea
H_{\rm eff} = \sum_{m}\sum_{n}\Omega^{n}_{m}\,\hat{\sigma}^+_{mn} \hat{\sigma}^-_{mn}~,
\label{H_eff}
\eea
where $\hat{\sigma}^{+(-)}_{mn}$ represents the creation (annihilation) operator for eigenmode $m$, where $m = 0, \pm 1, \ldots, \pm \left\lceil \frac{N_1 - 1}{2} \right\rceil$ and index $n\in \{1,2,3\}$ represents the three possible solutions for this layered concentric ring geometry. We note that the $m = \pm 1$ eigenmodes, which are central to the analysis in this article, are always degenerate regardless of whether $N_1$ is even or odd. As the lowest non-trivial winding modes, they possess the gentlest phase gradients - with a phase step of only $2\pi/N$ between adjacent sites and consequently offers the highest resilience to site disorder, since their long-wavelength modulation effectively averages over local perturbations in position and transition frequency. This property is shared with the optically bright $k = \pm 1$ exciton states of biological LH rings, such as the B850 complex of LH2, where matching eigenmode phase rotation to transition dipole geometry concentrates the collective oscillator strength into this degenerate pair~\cite{Schulten:jpcb:1997,vangrondelle:jpcb:1997}. An additional non-degenerate mode at $m = N_1/2$ appears only when $N_1$ is even. We mention this merely to note a distinction relevant when comparing with other ring geometries, such as photonic molecules~\cite{Bloch:np:2019}, but it does not participate in the analysis considered here.

This ring-configuration would always be immersed in a dissipative environment in reality, and the system-environment interaction~\cite{breuer2002theory} would generally influence the `pure' collective quantum optical signature greatly. However, for the purpose of this work, we completely neglect the dissipative part (under a reasonable approximation), as discussed in the following section.

\section{Closed Coupled-Dipole Model for LH2 Stacked Ring Geometry}
\label{ring-geometries}
We consider each BChl as a point optical dipole and a two-level system, considering only the lower excitation manifold~\cite{Schulten:jpcb:1997}. The transition frequencies are denoted as $\omega_{\nu_{0}}$ for the top ring $R_1$ and $\omega_{\nu}$ for the bottom ring $R_2$. In previous work~\cite{pal:njp:2025}, we discussed the partial validity of this Markovian master-equation approach in the context of LH2 complexes. We reiterate that while quantum optical models are instrumental, they may not fully capture the mechanisms underlying the near-unity efficiency of EET in biological systems. Moreover, the inter-molecular excitonic coupling~\cite{Cupellini:cms:2019} may not be correctly captured by $H_{\rm DD}$ alone. However, we remark that while this model is an approximation, it provides a valuable description of the `pure' electromagnetic layer of interacting dipoles that is submerged in the complex dissipative environment of the LH2 complex~\cite{pal:njp:2025}.

In recent theoretical works~\cite{kerber:prr:2026, nunner:arxiv:2026}, we have demonstrated that using an ultrashort (picosecond) Swing-UP of quantum EmitteR population (SUPER)~\cite{doris:prxq:2021} excitation scheme, employing two off-resonant, time-overlapped pulses, one can access the `pure' electromagnetic interactions of coupled emitters at deep-subwavelength separations, even amidst certain environmental decoherence, such as on-site position and frequency disorders. In reality, these environmental factors represent only a subset of the complex system-environment interactions in biological LH2. Given that the LH apparatus is typically suspended in the bacterial cell fluid, precise modeling of these complicated environmental fluctuations remains an open challenge relying solely on quantum optical frameworks~\cite{pal:njp:2025}. As biological LH2 complexes are extensively investigated via quantum chemistry~\cite{renger:pccp:2013,menucci:cr:2017, scholes:cr:2017}, we would like to clearly note that our intent is not to question these established frameworks (which actually provide the basis of studies on photosynthetic processes) nor to perform exhaustive quantum optical modeling to take into account the environmental effects~\cite{breuer2002theory}, as it lies beyond the scope of this study.

While our model defines the system as a `closed quantum-optical coupled-dipole system', this simplified approximation may also be valid in the limit of very high dissipation, where collective modes decay rapidly after initial excitation. Given that sunlight capture by the LH2 complex (diameter $\sim$ 6 nm) is an ultrafast single-photon absorption process, we restrict our analysis to the eigenmodes of the stacked-ring geometry, with different optical dipoles in different layers. Furthermore, to maintain scientific rigor, we refrain from characterizing the bright and dark nature of the collective optical modes, as dissipative signatures are neglected in this theoretical framework.

\begin{figure*}
\centering
\includegraphics[width=\textwidth]{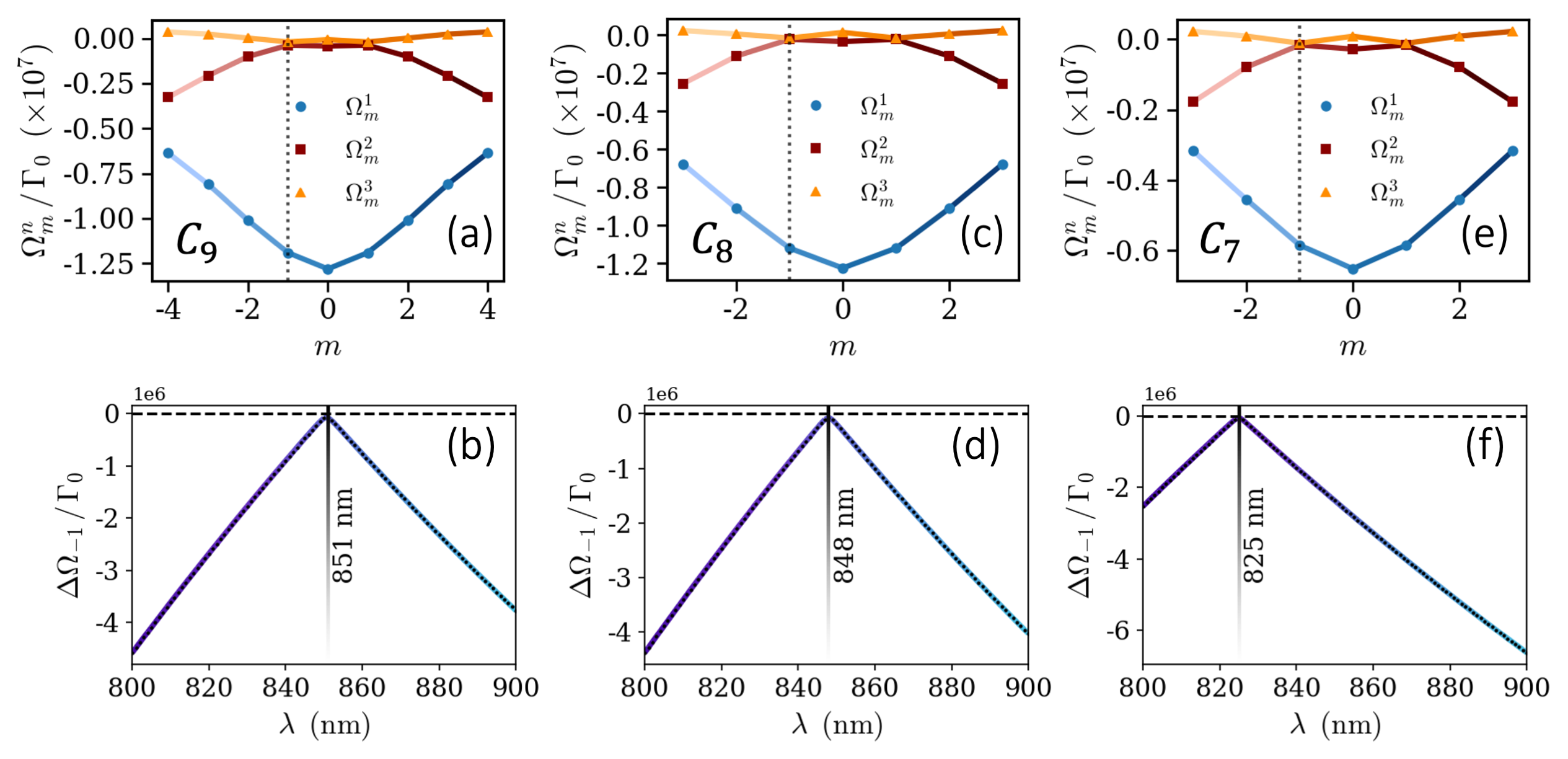}
\caption{Collective energy shift $(\Omega^n_m)$ for different angular momentum eigenmode values $m$ in LH2 ring geometries having ninefold ($C_9$) (a), eightfold ($C_8$) (c) and sevenfold ($C_7$) (e) symmetries. The black-dotted vertical line indicates the $m=-1$ mode. The energy band separation ($\Delta \Omega_m = (\Omega^2_{m} - \Omega^3_{m})/\Gamma_0$) for mode $m =-1$ with variable transition wavelength $\lambda$ for the optical dipoles of $R_2$ is plotted for $C_9$(b), $C_8$(d) and $C_7$(f) symmetries, while keeping all ring parameters unaltered. The minimum band separation, denoted by vertical gradient lines, indicates the photoactivity of the optical dipoles in ring $R_2$, where the ring $R_1$ has all 800 nm dipoles with $\Gamma_0 \sim 25.7$ MHz. Geometric parameters for the simulations are taken from Ref.~\cite{montemayor:jpcb:2018} for panels (a) and (b), and Ref.~\cite{pal:njp:2025} for panels (c)--(f).}
\label{LH2-bands}
\end{figure*}

In Ref.~\cite{pal:njp:2025}, we had observed that optimal EET, within the mentioned limits of our model, occurs primarily through eigenmodes characterized by a vanishing or minimal energy separation between specific collective energy bands. For the ninefold symmetry, this condition is met at the angular momentum modes $m = \pm 1$. In Figs.~\ref{LH2-bands}(a), (c), and (e), we plot the collective frequency shifts ($\Omega^n_m/\Gamma_0$) with different $m$'s for LH2 model rings with ninefold, eightfold, and sevenfold symmetries. Physically, when the energy separation between bands $\Omega^2_m$ and $\Omega^3_m$ approaches zero at a given transition wavelength $\lambda$, the corresponding collective eigenmodes of the top and bottom rings become near-degenerate. In this regime, the inter-layer hybridization between the $R_1$ and $R_2$ modes is maximized, facilitating efficient population transfer between the layers. The closer the collective bands, the stronger the mixing of top- and bottom-ring eigenstates~\cite{zener:prsl:1932, Novotny_Hecht_2006}, and consequently, leading to a more effective flow of excitation energy from $R_1$ to $R_2$. The wavelength at which this minimum separation occurs would identify the photoactive absorption optimum of the lower ring $R_2$. For simulation, the geometric parameters are taken from Refs.~\cite{montemayor:jpcb:2018} (ninefold) and \cite{pal:njp:2025} (seven and eightfold). The spontaneous emission rate for an 800 nm dipole is $\Gamma_0 \sim$ 25.7 MHz.

Our numerical results indicate that the energy bands $\Omega^2_m, \Omega^3_m$ consistently overlap at $m = \pm 1$ across all three deep-subwavelength geometries [see Figs.~\ref{LH2-bands}(a), (c), and (e)] at a specific wavelength for the photoactive dipoles in the lower ring $R_2$. This behavior is a direct consequence of the specific geometric dimensions and dipole orientations inherent in these ring configurations. By scanning the transition wavelength $\lambda$ for the lower dense ring $R_2$, we find that the minimum separation $|\Delta\Omega_{-1}|$ occurs at 851 nm for $C_9$ (close to reported 850 nm~\cite{kuhlbrant:structure:1995, kohler_2006}), at 848 nm for $C_8$ (reported 850 nm~\cite{koepke:structure:1996,schulten:pnas:1998}) and at 825 nm for $C_7$ (close to 828 nm reported in~\cite{qian:sa:2021,cupellini:pr:2023}). Indeed, it seems {\it remarkable} that a closed quantum optical model, despite its simplifications, effectively captures the photoactive absorption peaks of the LH2 lower ring, in agreement with previous studies.

Note that the onsite disorders could enhance the bandwidth of EET and light absorption, which is not considered here, and is also not important for the purpose of this work. Also note that as the fold number decreases (from $C_9$ to $C_7$), the decreased density of the lower ring results in a blue shift of the B850 band [Figs.~\ref{LH2-bands}(b), (d), and (f)], as previously also noted in Ref.~\cite{pal:njp:2025}. Physically, as the fold number decreases at fixed ring diameter, the number of BChls in the lower ring drops. Additionally, the increased inter-BChl separations weaken the near-field dipole--dipole couplings $\Omega_{ij}$ [Eq.~(\ref{omegaij})], which scale as $1/r_{ij}^3$ in the deep-subwavelength limit ($r_{ij} \ll \lambda$). Moreover, for the $m = \pm 1$ collective eigenmodes, the phase increment between adjacent BChls is $2\pi/N$, where $N$ is the fold number; as $N$ decreases, this phase step increases, altering how the individual dipole--dipole couplings sum into the collective eigenvalue and further shifting the wavelength at which the resonance condition for inter-layer EET is satisfied. This reduced collective coupling diminishes the red-shift of the lower band relative to the monomer transition frequency, further shifting the predicted photoactive wavelength toward shorter values and away from the peak of the available solar spectrum. For completeness, Appendix~\ref{apen-eet} evaluates the maximum EET achievable at optimized values of $\lambda$ for all three symmetries. However, we reiterate that this closed quantum optical framework is inherently limited~\cite{pal:njp:2025} in its ability to capture the near-unity efficiency of the biological LH complexes.

\section{Investigating Sub-Sevenfold Symmetries in LH2 Ring Scaffolds via the Coupled-Dipole Model}
\label{sub-sevenfold}

The photosynthetic apparatus of purple photosynthetic bacteria, weakly illuminated by sunlight, receives approximately 10 photons per second per chlorophyll molecule~\cite{blankenship2002molecular}. This situation can be described as a rain of incoming photons hitting the molecule, or simultaneously as wave interactions, given that the wavelength is much larger than the biological ring size. From the perspective of collective light--matter interaction, the light absorption process in the LH2 apparatus can be visualized as an antenna with a specific collective optical dipole (whose radiative properties are not characterized here) and a corresponding absorption cross-section. For a $C_9$ geometry, the total number of photons absorbed per 0.1 ms ($10\times\pi r^2_1\times 10^{-4}$) is $\sim$  3. The excitation transfer time from top B800 band $\rightarrow$ lower B850 band is around 0.6 ps~\cite{FLEMING:sc:1997} and the lifetime of one BChl molecule is about 39 ns (corresponding to $\Gamma_0 \sim$ 25.7 MHz). It is important to note that collective modes exhibit modified lifetimes. In previous work~\cite{pal:njp:2025}, we observed that the modified decay rate for the $m = \pm 1$ modes of the B800 ring in $C_9$ LH2 is $\sim 10^{4.4}~\Gamma_0$, i.e., 1.5 ps, same as noted in Ref.~\cite{vanGrondelle:JPCB:1999} at 4 K (noting that lifetimes typically increase as temperature decreases).

These values suggest that photo-absorption and the subsequent inter-layer transfer $R_1\rightarrow R_2$ is quite an optimal process, and the quantum optical description may hold some validity. Furthermore, previous studies have indicated that entanglement likely plays no key role in the excitation transport mechanism of photosynthesis~\cite{Plenio_2008, briegel:rsca:2012}. Our closed quantum optical model is considered valid only within a narrow temporal window (on the order of a fraction of a picosecond) shorter than the B800$\rightarrow$B850 EET time (see Appendix~\ref{apen-eet} for quantitative details). According to our description, the collective mode must be repeatedly excited by impinging photons to drive the EET process, which probably aligns with the experimentally observed coherence time of $\sim$ 400 fs in photosynthesis~\cite{vanhulst:science:2013}.

\begin{table}
\caption{\label{tab1}
\textit{Assumed} geometric parameters for a possible sixfold ($C_6$) LH2 ring conformation. The dipole orientations are considered to be the same as those of \textit{Rhodoblastus acidophilus} with $C_9$ symmetry and taken from~\cite{montemayor:jpcb:2018}. The top ring is $R_1$ and has $N_1 = 9$ emitters and $R_2$ is the bottom ring formed of two sub-rings: ${R_{2_a}}$ and ${R_{2_b}}$, each having $N_{2_{a(b)}} = 9$ emitters (see Fig.~\ref{schematic}(a)). The ring radii are $r_1$, $r_{2_a}, r_{2_b}$ for rings $R_1, R_{2_a}$ and $R_{2_b}$, respectively. The dipole moments are 6.48 D, 6.41 D, and 6.3 D for rings $R_1$, $R_{2_a}$, and $R_{2_b}$, respectively. The vertical layer separation is $Z_1$ (see Ref.~\cite{pal:njp:2025} for a comprehensive schematic of $C_9$ stacked ring scaffold).}
\begin{ruledtabular}
\begin{tabular}{ccc}
Ring size & {Ring rotation} &  {Dipole orientations}\\
($r_i$,$Z_1$) (\AA)&  $(\nu_i)$ (deg) &  $(\theta_i, \phi_i)$ (deg) \\
\hline
${r_1}$ \hskip 0.5cm 28.5 &  $\nu_1$ \hskip 0.3cm $0^{\circ}$ &  $\theta_{1},\phi_1$ \hskip 0.4cm $98.2^{\circ}, 63.7^{\circ}$\\
$Z_{1}$ \hskip 0.5cm 18 &  & \\
\hline
$r_{2_a}$ \hskip 0.5cm 25.4  &   $\nu_{2_a}$ \hskip 0.3cm $-15^{\circ}$ &  $\theta_{2_a},\phi_{2_a}$ \hskip 0.2cm $96.5^{\circ}, -106.6^{\circ}$\\
$r_{2_b}$ \hskip 0.5cm 26 &   $\nu_{2_b}$ \hskip 0.4cm $15^{\circ}$ &  $\theta_{2_b},\phi_{2_b}$ \hskip 0.4cm $97.3^{\circ}, 60.0^{\circ}$ \\
\end{tabular}
\end{ruledtabular}
\end{table}

Our approximation suggests that the nature of the collective modes, specifically, whether the underlying collective quantum states in the weakly driven single-excitation subspace are symmetric or antisymmetric, does not necessarily imply that biological LH geometries rely on specific bright or dark states to execute EET; at this stage, this remains rather inconclusive. Furthermore, for efficient photon absorption, the cross-section of ring $R_1$ must be sufficiently large to optimize the use of the available solar spectrum. To maintain an ideal absorption cross-section, we have kept the ring diameter for the hypothetical $C_6$ geometry similar to that of the $C_7$ scaffold~\cite{pal:njp:2025, qian:sa:2021, cupellini:pr:2023} (see Table~\ref{tab1} for geometric parameters). The absorption cross-section of the LH2 antenna scales with its geometric area, i.e., $\pi r^2_1$; reducing the ring diameter to accommodate only six subunits at the same inter-BChl spacing would decrease the cross-section substantially, lowering the photon capture rate in the already photon-limited environment of purple bacteria. Crucially, a smaller ring would also alter the inter-dipole separations entering $\Omega_{ij}$ [Eq.~(\ref{omegaij})], and could lead to a shift of the predicted photoactive wavelength. Our choice to preserve the ring diameter therefore represents the most favorable scenario for the hypothetical $C_6$ geometry, as it grants $C_6$ the largest possible antenna area and the mildest possible spectral shift. We note that a more precise estimate would require detailed knowledge of the ring parameters, including the dipole orientations, which is unfortunately a major limitation. Despite this limitation, the best-case configuration nevertheless yields an absorption peak that compromises light-harvesting efficiency.

It is important to note that purple photosynthetic bacteria typically inhabit anaerobic layers in aquatic environments where the available solar flux is spectrally shaped by multiple filtering mechanisms before it reaches the bacteria. The solar spectrum that arrives at the depth of purple bacteria has been filtered by the chlorophylls of oxygenic phototrophs - located in the upper layers of the water column, which effectively remove most of the blue and red light~\cite{blankenship2002molecular, kohler_2006}. Purple bacteria must therefore harvest wavelengths above this range. In practice, the LH2 scaffolds contain both BChls (the focus of our study) and carotenoids (not considered here): carotenoids harvest shorter wavelengths~\cite{polivka:cr:2004}, while BChls harvest the near-infrared portion. To sustain their metabolism, these organisms must capture all usable photons within their spectral niche. The LH2 architecture is optimized for precisely this window: the B800 band of the upper ring $R_1$ absorbs at $\sim$800~nm, and the excitation is subsequently transferred to the lower ring $R_2$, whose B850 band extends the spectral coverage toward the long-wavelength edge of the available flux. The LH2 geometries observed in nature place their lower-ring photoactive absorption well within this niche, at 851~nm ($C_9$), 848~nm ($C_8$), and 825~nm ($C_7$), the last already lying at the edge of optimal harvesting. Our predicted 814~nm for the hypothetical $C_6$ scaffold falls further below this favorable range, into the spectral gap between the B800 absorption and the peak of the bacterial photonic environment, rather than extending coverage toward longer wavelengths where the available flux is concentrated. This spectral mismatch arising from fewer BChls in the lower ring would further compromise the light-harvesting capacity in the already photon-limited environment of purple bacteria~\cite{kohler_2006}. While one could in principle design a smaller ring with sixfold ($C_6$) symmetry - we note that there is no physical constraint, in our understanding and opinion, but nature has not designed them, as not reported in the literature. On this note, the resulting attenuation of the absorption cross-section would yield a sub-optimal light-harvesting capacity given the specific local photonic environment of purple bacteria~\cite{blankenship2002molecular, kohler_2006}.

\begin{figure}
\centering
\includegraphics[width=\columnwidth]{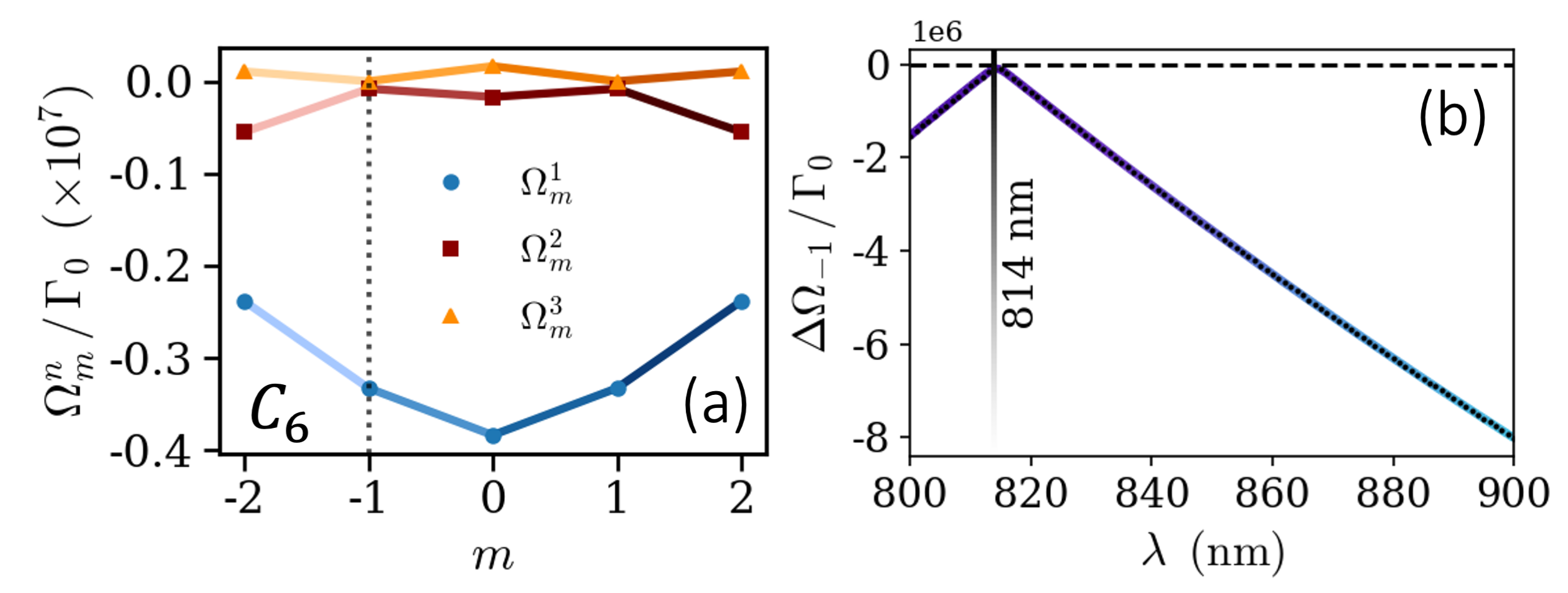}
\caption{(a) Collective energy shift $(\Omega^n_m/\Gamma_0)$ for an \textit{assumed} sixfold $C_6$ symmetry with angular momentum quantum number $m$. Table~\ref{tab1} illustrates the geometric parameters in use for this \textit{assumed} geometry. (b) The energy band separation at $m = -1$ (dotted black line in panel (a)), i.e., $\Delta \Omega_{-1} = (\Omega^2_{-1} - \Omega^3_{-1})/\Gamma_0$, for variable transition wavelength $\lambda$ (corresponding to the optical dipole of bottom ring $R_2$) under $C_6$ symmetry with all other parameters fixed. The smallest band separation is denoted by the vertical line in panel (b). The top ring $R_1$ has all 800 nm dipoles, as in the earlier cases displayed in Fig.~\ref{LH2-bands}.}
\label{c6}
\end{figure}

In Fig.~\ref{c6}, panel (a) shows the collective energy shift for the $C_6$ model ring scaffold. The minimum energy separation $|\Delta\Omega_m|$ of bands $\Omega^2_m, \Omega^3_m$ at the angular momentum mode $m = -1$ suggests that the photoactivity of a sixfold geometry occurs at approximately 814~nm. We note that any change in ring size would cause an additional spectral shift of the predicted wavelength. As discussed above, this blue shift places the $C_6$ absorption outside the optimal spectral window, thereby reducing overall efficiency. If our simple quantum optical analysis is able to catch a little bit of Nature's thinking, then perhaps {\it her} supreme optimization for most efficient and resilient conformations in use for light-harvesting mechanisms could raise a legitimate question on the existence of the sixfold $C_6$ LH2 scaffolds. If it remains true, similar principles may account for the observed absence of all sub-sevenfold symmetries in biological LH2 ring scaffolds. We emphasize that the analysis presented here offers a quantum-optical perspective on the absence of sub-sevenfold symmetries, and that our reasoning is inductive in nature: having validated the closed coupled-dipole model against three experimentally known LH2 symmetries ($C_9$, $C_8$, and $C_7$), we apply the same framework to the hypothetical $C_6$ geometry. Our analysis does not derive, from first principles, a fundamental prohibition of sub-sevenfold symmetries; rather, it yields a numerically grounded and experimentally falsifiable prediction that such geometries would be spectrally inefficient for light-harvesting within the photonic environment of purple bacteria.

\section{Conclusions and Outlook}
\label{conclusion}
By utilizing a closed coupled-dipole model under appropriate approximations, relying on eigenanalysis and the specific photonic environment of purple bacteria, we theoretically demonstrate that sub-sevenfold symmetries are likely absent in LH2 ring scaffolds because such geometries lack maximal efficiency. Within our model, this inefficiency originates from the reduced BChl density at fixed ring size, which weakens the collective dipole-dipole couplings and diminishes the red-shift of the lower-ring absorption band, displacing it from the optimal spectral window of the bacterial photonic environment. Our framework assumes that this quantum-optical description is valid only within an extremely short temporal window, significantly shorter than the interlayer EET time. Consequently, within this model, the continuation of the EET process requires repetitive excitation.

A limitation of our approach is that it remains inconclusive regarding whether bright or dark optical modes are responsible for energy transport. Furthermore, it also does not indicate aspects of entanglement in the energy transfer of LH complexes, as also noted earlier in Refs.~\cite{Plenio_2008,briegel:rsca:2012, sarovar:np:2010}. Note that these are not the focus of this study. Our investigation suggests that a purely quantum-optical model or a Markovian description~\cite{pal:njp:2025} may not fully capture the complexities of the photosynthetic EET mechanism. The quantum-optical Born--Markov treatment in Ref.~\cite{pal:njp:2025} is useful for isolating collective dipole-mediated dynamics, but it is an idealized description of biological LH2. It does not explicitly represent the full pigment--protein environment, including structured vibronic coupling and carotenoid-mediated pathways, and it treats environmental effects in a simplified way. Such factors can influence dephasing and EET pathways~\cite{fleming:pnas:2009,Dubi:acso:2023}. For completeness, it is worth noting that various natural systems exhibit sub-sevenfold symmetries - such as the sixfold symmetry of snowflakes~\cite{Libbrecht:rpp:2005}. At the right temperature and humidity, water molecules nucleate as hexagonal crystals on dust particles in the atmosphere, and as each crystal traverses toward the ground, subsequent growth preserves this symmetry. One also finds, for example, fivefold symmetry in starfish (arising from developmental constraints and structural considerations) and various floral symmetries (shaped by pollination ecology and genetics)~\cite{JABBOUR:crb:2009}. These patterns arise from different physical, developmental, and evolutionary constraints and are not mechanistic analogs of LH2 ring geometry. In LH2, the relevant structural questions are how the fold symmetry arranges the pigments and how that arrangement, together with membrane packing, influences excitonic coupling and inter-ring EET - a spectral selectivity that prior symmetry-dependent EET models~\cite{cao:pnas:2013,Kassal:jpcl:2016} did not explicitly address.

Circular patterns also emerge in various other physical systems, including planar ion crystals~\cite{okada:pra:2010, bollinger:nature:2012, Freericks:EPJqt:2015, richerme:prl:2021}, bilayer ion crystals~\cite{shankar:prx:2024}, and dipolar~\cite{ferlaino:nature:2021, bisset:prl:2022, stringari:nrp:2023} or layered supersolids~\cite{bohn:pra:2024}. Whether and to what extent these systems are relevant for studying transport mechanisms~\cite{roos:prl:2019} remains an intriguing open question. Additionally, it would be worthwhile to investigate how naturally occurring frequency shifts in subwavelength arrangements, for instance, in atomic arrays~\cite{hofer:prxq:2025}, Rydberg atoms~\cite{zeiher:np:2023}, and quantum dots~\cite{trotta:prl:2018, Dalacu_2019}, as well as semiconductor microcavity photonic molecules with discrete rotational symmetry and tunable angular momentum modes~\cite{Bloch:np:2019}, could be useful to study excitation transfer or, more broadly, light-harvesting mechanisms. Our investigation also contributes to certain domains of quantum biology~\cite{van_grondelle:jrsci:2018, scholes:pnas:2026}, where suitable quantum cooperative models, under reasonable approximations, can offer valuable insights. Specifically, our effort to reduce the vast complexity of biological reality to an elegant, accessible framework to elucidate certain fundamental aspects of living entities represents an enlightening research direction to pursue.

\section*{Data availability}
The simulations in this manuscript were performed utilizing the QuantumOptics.jl~\cite{KRAMER2018109} and CollectiveSpins.jl~\cite{collectivespinsjl} frameworks in the Julia programming language. The plots were prepared using the Matplotlib~\cite{matplotlib} library. The data that support the findings of this manuscript are
openly available at Ref.~\cite{pal_zenodo_2026}.

\begin{acknowledgments}
A.P. gratefully acknowledges Klemens Hammerer for kindly reading the manuscript and providing valuable comments. She extends her sincere thanks to Hans J. Briegel for insightful conversations and illuminating suggestions concerning her broader research interests. A.P. is particularly grateful to Helmut Ritsch for a valuable discussion and for generously sharing the Julia code for simulating collective dipole dynamics, which was utilized in parts of the numerical framework of this work. She also thanks Isobel C. Bicket for a helpful discussion on the collective dipole description. A.P. is grateful to Lorenzo Cupellini for valuable comments from the quantum chemistry perspective, some of which have been incorporated into the present manuscript. She also gratefully acknowledges the comments of Erik M. Gauger and the anonymous referees of her previous work~\cite{pal:njp:2025}, as their insights have informed aspects of the present study. She further thanks Mikhail O. Nestoklon and Moritz Cygorek for useful discussions and helpful comments. A.P. also gratefully acknowledges the opportunity to participate in the `Workshop on Long-Range Interactions in Quantum Systems 2026' (\'Ecole de Physique des Houches), `Quantum Optics 2026' (Universit\"atszentrum Obergurgl), `NanoLight 2026' (Centro de Ciencias de Benasque Pedro Pascual), and `Quantum Machine Learning' (Bad Honnef Physics School), as the scientific exchanges during these visits contributed to shaping certain discussions presented in this work. Finally, she extends her warm thanks to Maria Moreno-Cardoner and Raphael Holzinger for general discussions during her earlier research on ring systems~\cite{pal:njp:2025}. This research was funded in whole or in part by the Austrian Science Fund (FWF) [grant DOI: 10.55776/ESP682].
\end{acknowledgments}

\appendix
\section{Quantitative Temporal Windows for inter-layer EET in Model LH2 Ring Geometries}
\label{apen-eet}
We consider a particular eigenmode $m'$ from the top ring $R_1$ (specifically $m'=\pm 1$ in this study) and estimate the total population transferred to the bottom ring $R_2$ denoted as $P_{R_2}^{m'}(t)$, which exhibits finite temporal evolution, as follows,
\bea
P_{R_2}^{m'}(t) = \sum_{k \in R_{2}} \langle \hat{\sigma}^{+}_{k} \hat{\sigma}^{-}_{k} \rangle_{\{m' \in R_1\}} ~.
\eea
Within a very small time interval - much less than the interlayer EET time - the maximum population transfer from top to the bottom ring, ${\rm Max}[P_{R_2}^{m'}(t)]$ yields values of approximately 0.35 ($\sim$ 0.3 ps), 0.37 ($\sim$ 0.3 ps), 0.89 ($\sim$ 0.39 ps), and 0.89 ($\sim$ 0.47 ps) for ninefold, eightfold, sevenfold, and sixfold rotational symmetries, respectively, with the corresponding temporal windows noted in parentheses. In all instances, the estimated maximum transfer remains below unity. This is expected, as our modeling is insufficient to capture the highly efficient EET observed in natural photosynthesis~\cite{pal:njp:2025}, and is also not the main focus of this investigation.

\bibliography{draftbib}

\end{document}